\documentclass[12pt]{article}

\usepackage{amsmath}
\usepackage{amssymb}
\usepackage{epsfig}
\usepackage{graphicx}
\usepackage{cite}
\usepackage{multirow}
\usepackage{longtable}
\usepackage{rotating}
\usepackage{bbm}

\allowdisplaybreaks[4] 

\makeatletter
\renewcommand{\fnum@table}{\textbf{\tablename~\thetable}}
\renewcommand{\fnum@figure}{\textbf{\figurename~\thefigure}}
\makeatother

\newcounter{myenumi}

\renewcommand{\themyenumi}{\roman{myenumi}}
{\end{list}}

\newlength{\myem}
\newcounter{mysubequation}[equation]

\makeatletter
\renewcommand{\section}{\@startsection{section}{1}{0em}{-\baselineskip}%
{\baselineskip}{\normalfont\large\bfseries}}
\renewcommand{\subsection}%
{\@startsection{subsection}{2}{0em}{-0.7\baselineskip}%
{0.7\baselineskip}{\normalfont\bfseries}}
\makeatother

\newcommand{\ie}{{\it i.e.}}

\newcommand{\eg}{{\it e.g.}}

\newcommand{\cf}{{\it cf.}}

\newcommand{\fig}{Fig.}

\newcommand{\Ref}{Ref.}
\newcommand{\Refs}{Refs.}

\newcommand{\Tab}{Tab.}

\newcommand{\stheta}{\sin^22\theta_{13}}
\newcommand{\deltacp}{\delta_\mathrm{CP}}
\newcommand{\ldm}{\Delta m_{31}^2}

\newcommand{\figu}[1]{\fig~\ref{fig:#1}}

\newcommand{\bi}{\begin{itemize}}
\newcommand{\ei}{\end{itemize}}

\newcommand{\brli}{($^8$B,$^8$Li)}
\newcommand{\nehe}{($^{18}$Ne, $^6$He)}

\begin{document}

\begin{titlepage}

\renewcommand{\thefootnote}{\alph{footnote}}

\vspace*{-3.cm}
\begin{flushright}
\end{flushright}


\renewcommand{\thefootnote}{\fnsymbol{footnote}}
\setcounter{footnote}{-1}

{\begin{center}
{\large\bf Minimal Neutrino Beta Beam for Large $\boldsymbol{\theta_{13}}$
} \end{center}}
\renewcommand{\thefootnote}{\alph{footnote}}

\vspace*{.8cm}
\vspace*{.3cm}
{\begin{center} {\large{\sc
                Walter~Winter\footnote[1]{\makebox[1.cm]{Email:}
                winter@physik.uni-wuerzburg.de}
                }}
\end{center}}
\vspace*{0cm}
{\it
\begin{center}

       Institut f{\"u}r Theoretische Physik und Astrophysik, Universit{\"a}t W{\"u}rzburg, \\
       D-97074 W{\"u}rzburg, Germany

\end{center}}

\vspace*{1.5cm}

{\Large \bf
\begin{center} Abstract \end{center}  }

We discuss the minimum requirements for a
neutrino beta beam if $\theta_{13}$ is discovered by an upcoming reactor experiment, such as Double Chooz or
Daya Bay. We require that both neutrino mass hierarchy and leptonic CP violation can be measured to competitive precisions with a single-baseline experiment in the entire remaining $\theta_{13}$ range.
We find that for very high isotope production rates, such as they might be possible using a production ring,
a \brli\ beta beam with a $\gamma$ as low as 60 could already be sufficient to perform all of these measurements. 
If only the often used nominal source luminosities can be achieved, for example, a \nehe\ beta beam from Fermilab to a possibly existing water Cherenkov detector at Homestake with $\gamma \sim 190-350$ (depending on the Double Chooz best-fit)  could outperform practically any other beam technology including wide-band beam and neutrino factory.

\vspace*{.5cm}

\end{titlepage}

\newpage

\renewcommand{\thefootnote}{\arabic{footnote}}
\setcounter{footnote}{0}

\section{Introduction}

In elementary particle physics, the main motivation to push to higher energies
is the search for physics beyond the standard model. So far, there has been
some evidence for such physics, such as the presence of dark matter or
the observation of neutrino oscillations requiring a non-vanishing neutrino mass.
It is therefore important to understand these indications for new physics
very carefully.
In neutrino oscillation physics, the so-called solar and atmospheric
oscillation parameters have been measured to high precisions, see, \eg,
\Ref~\cite{GonzalezGarcia:2007ib}. However, 
we only have an upper bound for the reactor mixing angle $\theta_{13}$, and we
do not know the mass ordering (normal or inverted), the absolute neutrino mass scale,
and the nature of neutrino mass (Dirac or Majorana). Furthermore,
there may be (Dirac) CP violation in the lepton sector, which is described by $\deltacp$.
For example, a detection of leptonic CP violation together with a $0\nu\beta\beta$
signal, which indicates that neutrinos are mostly Majorana particles, will 
motivate leptogenesis as a mechanism to produce the dominance of matter over antimatter in the early
universe. In addition, a determination of $\theta_{13}$ and the mass ordering will help our understanding 
of stellar evolution~\cite{Dighe:2007ks}, and these parameters turn out to be
excellent discriminators for neutrino mass models including grand unified theories~\cite{Albright:2006cw}.
Therefore, future neutrino oscillation experiments may use
high energy neutrino beams over long distances to study the remaining unknown oscillation parameters
$\theta_{13}$, $\mathrm{sgn}(\ldm)$ (which we call mass hierarchy), and $\deltacp$, while
nuclear physics experiments, such as $0\nu\beta\beta$ decay and tritium endpoint measurements, will probe
absolute neutrino mass scale and the nature of the neutrino mass.
An early determination of $\theta_{13}$ might already be possible by upcoming reactor
experiments, such as Double Chooz or Daya Bay~\cite{Ardellier:2006mn,Guo:2007ug}.

For the beam experiments, there are, in principle, different approaches depending
on the magnitude of $\theta_{13}$. Superbeams, such as T2K or NO$\nu$A~\cite{Itow:2001ee,Ayres:2004js},
or upgrades thereof, are based on neutrino production
by pion or kaon decays using a high intensity proton beam on a target. This technique
works especially well for large $\theta_{13}$, where the backgrounds are of little relevance. Potential future neutrino factories~\cite{Geer:1998iz,Apollonio:2002en,ids} use neutrino production by muon decays. They are discovery machines with an excellent reach in $\theta_{13}$. A beam production technique, which is intimately connected to
nuclear physics, is used by so-called beta beams~\cite{Zucchelli:2002sa,Mezzetto:2003ub,Autin:2002ms,Bouchez:2003fy,Lindroos:2003kp}. For these beams, unstable nuclei, such as from
the pairs \brli\ or \nehe , decay in straight sections of a storage ring to produce
an electron flavor-clean $(\nu_e,\bar{\nu}_e)$ neutrino beam. 
There are two key components
for such an experiment: A high intensity ion source, characterized by the number of produced ions per
time frame,  and a sufficiently large 
accelerator to boost the ions to higher energies, characterized  by the boost factor $\gamma$.
For the source, several approaches have been studied in the literature. 
For example, the ISOL (Isotope Separation On-Line) technique~\cite{EURISOL} could also be used for 
a wider range of nuclear physics, which means
that there will be a lot of synergies in the neutrino oscillation and nuclear physics programs.  
In addition, the direct production method with
a storage ring might lead to even higher source luminosities than originally anticipated, which has been proposed for \brli ~\cite{Mori:2005zz,Rubbia:2006pi,Rubbia:2006zv}.
For the accelerator, either an existing machine might be used (such as the CERN-SPS or the Tevatron), or a new one might be built. The difference in the \brli\ and \nehe\
ion pairs is their endpoint energies $E_0$: Since the peak of the spectrum is approximately
given by $E_\nu \sim E_0 \cdot \gamma$, a lower $\gamma$ might be sufficient if a higher $E_0$ can be used,  such as in \brli\ compared to \nehe . However, a lower $\gamma$ means worse beam collimation, which leads to lower event rates. The interplay between isotope pair, $\gamma$, and ion source luminosity is therefore non-trivial~\cite{Agarwalla:2008gf}.

Beta beams have been studied in specific scenarios from low to very high $\gamma$'s~\cite{BurguetCastell:2003vv,BurguetCastell:2005pa,Agarwalla:2005we,Campagne:2006yx,Donini:2006dx,Donini:2006tt,Agarwalla:2006vf,Agarwalla:2007ai,Coloma:2007nn,Jansson:2007nm,Meloni:2008it,Agarwalla:2008ti}, and there has been a green-field optimization to push the sensitivities for small $\theta_{13}$~\cite{Huber:2005jk,Agarwalla:2008gf}. 
In almost all cases, the luminosities and $\gamma$'s are more or less chosen arbitrarily from the physics point of view, whereas they are rather determined by technical boundary conditions in many cases.
However, in the context to alternative superbeams and neutrino factories, the neutrino oscillation physics case of a beta beam might actually be defined by a (large) $\theta_{13}$ signal of the the upcoming beam or reactor experiments, such as Double Chooz or Daya Bay. In this work, we therefore discuss the {\em minimal} requirements for a beta beam to outperform any of its alternatives if Double Chooz finds $\theta_{13}$.
 For example, it is yet unclear if the $\gamma \simeq 350$ in \Ref~\cite{BurguetCastell:2005pa}, which has an excellent  performance, is really the minimal allowable setup.
Compared to the small $\theta_{13}$ case, in which one optimizes for $\theta_{13}$ reaches as good as possible, the definition of the minimum wish list from the physics point of view is rather straightforward:
\begin{itemize}
\item
 $5\sigma$ independent confirmation of $\stheta>0$
\item
 $3\sigma$ determination of the mass hierarchy (MH) for {\em any} (true) $\deltacp$
\item
 $3\sigma$ establishment of CP violation (CPV) for 80\% of all (true) $\deltacp$
\end{itemize}
in the {\em entire remaining allowed $\theta_{13}$ range}. Note that
we do not know the (true) $\deltacp$ which nature has implemented, which significantly affects the 
sensitivities. Therefore, we follow a low risk strategy and postulate that our experiment works
for any value of this parameter. The only exception  is the fraction of $\deltacp$ for CPV: 
Since $\deltacp =0$ and $\pi$ 
are both CP conserving, one cannot measure CPV for any $\deltacp$. The arbitrarily chosen fraction 
80\% can be motivated
by a competitive precision compared to a neutrino factory~\cite{ids}, or, as we will see
later, compared to many other facilities.

But what means ``minimal'' in terms of technical effort for a beta beam? Certainly, minimal refers to using only
one baseline. For a given detector, minimal refers to a yet-to-be-defined product between accelerator cost ($\propto \gamma$) and ion source intensity. 
We study the minimal effort for the above measurements in terms of this product quantitatively, and we discuss the dependence on the isotopes and detector technology used.

\begin{table}
\begin{center}
\begin{tabular}{lllr}
\hline
$\stheta$ best-fit & $90\%$ CL range & $3\sigma$ range & Zero excl. at \\
\hline
0.04 & 0.019 - 0.063 & 0.002 - 0.082 & $3.2 \sigma$ \\ 
0.08 & 0.060 - 0.102 & 0.043 - 0.121 & $6.4 \sigma$ \\
0.12 & 0.100 - 0.142 & $\ge$ 0.084 & $9.7 \sigma$ \\
\hline
\end{tabular}
\end{center}
\caption{\label{tab:dchooz} Several best-fit values for Double Chooz (first column), and the allowed range for $\stheta$ (second, third columns). The fourth column gives the
exclusion power of $\stheta=0$. Simulation from \Ref~\cite{Huber:2006vr} for 3 years of far detector
operation and 1.5 years of near detector operation.}
\end{table}

\section{Method}

We assume that Double Chooz finds $\stheta$, and we require that the above conditions are met for {\em any} $\stheta$ within the 90\% CL allowed region of Double Chooz (\cf, \Tab~\ref{tab:dchooz} for several simulated best-fit values). Note that the current bound on $\stheta$ is $0.157$ at $3\sigma$~\cite{GonzalezGarcia:2007ib}.
We use \nehe\ and \brli\ as possible isotope pairs, with $1.1 \cdot 10^{18}$ ($\nu_e$) and $2.9 \cdot 10^{18}$ ($\bar{\nu}_e$) useful ion decays per year, respectively, which are the nominal isotope decay rates often chosen in the literature~\cite{Terranova:2004hu}. For the sake of simplicity, we operate each ion at the {\em same} $\gamma$ for neutrinos and antineutrinos for five years, \ie, we assume a total running time of ten years. As detectors, we use a 100~kt Totally Active Scintillating Detector (TASD), which could be replaced by a liquid argon detector for a similar performance, and a 500~kt water Cherenkov detector (WC);
see \Refs~\cite{Huber:2002mx,Huber:2005jk} for simulation details. Note that for large $\stheta$, the cuts for both detectors should account for high efficiencies rather than low backgrounds, because in this limit, backgrounds are less relevant.
We use $\gamma \lesssim 500$ as the allowed $\gamma$ range, unless \brli\ is combined with the WC detector, where we use $\gamma \lesssim 150$ to avoid an un-predictive detector behavior due to too large neutrino energies.
Our simulations use the GLoBES software~\cite{Huber:2004ka,Huber:2007ji} with the current best-fit values and solar oscillation parameter uncertainties from \Ref~\cite{GonzalezGarcia:2007ib}, as well as a 2\% error on the matter density profile. For the sake of simplicity, we use a normal simulated mass hierarchy. The uncertainty on the atmospheric oscillation parameters is simulated by the inclusion of 10 years of T2K disappearance data.

In some cases, we will discuss our results as a function of the {\em luminosity scaling factor} $\mathcal{L}$, which scales the product of useful ion decays per year $\times$ running time $\times$ detector mass $\times$ detection efficiency. Thus, $\mathcal{L}=1$ corresponds to our nominal luminosity, whereas $\mathcal{L}=5$ corresponds to, for example, scaling up the detector mass by a factor of two and the source luminosity by a factor of $2.5$.

\begin{figure}[t]
\begin{center}
\includegraphics[width=10cm]{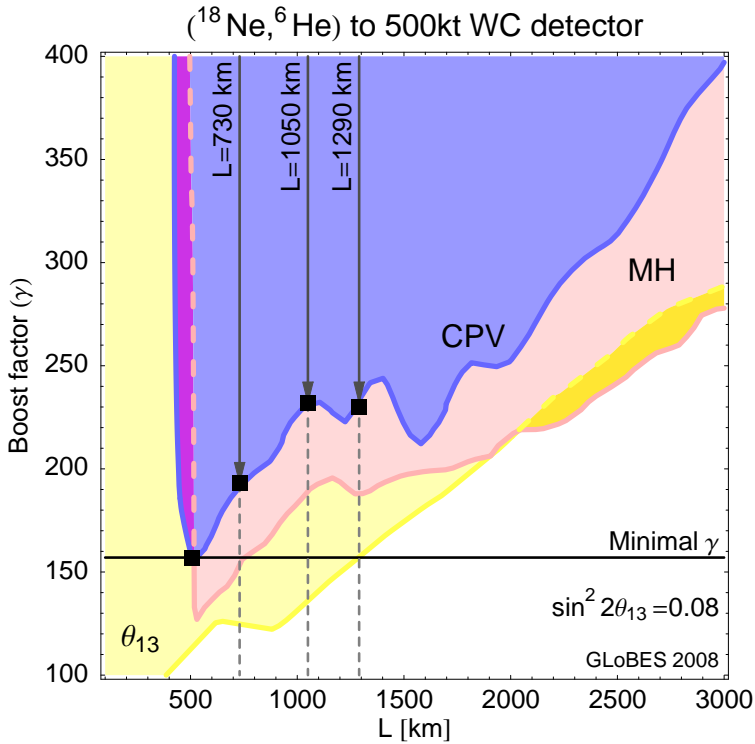}
\end{center}
\caption{\label{fig:lgamma1}Discovery of $\theta_{13}$ (dark/blue), a normal MH (medium gray/red),
and CPV (light gray/yellow) as a function of baseline $L$ and boost factor $\gamma$.  Sensitivity is given within the shaded regions at the $5\sigma$ CL for $\theta_{13}$ (for all values of true $\deltacp$), at the $3\sigma$ CL for the MH (for all values of true $\deltacp$), and at the $3\sigma$ CL for CPV (for at least 80\% of all possible true $\deltacp$). The minimal possible $\gamma$, as well as the minimal~$\gamma$'s for specific baselines, are marked. The figure is computed for the WC detector and \nehe , and $\stheta=0.08$ (best-fit) from \Tab~\ref{tab:dchooz}.}
\end{figure}

\section{Results}

For a given $\mathcal{L}$, isotope pair, and detector, the minimal effort is determined by the minimal~$\gamma$ for {\em any} baseline $L$. Therefore, we need to perform an optimization in the $L$-$\gamma$ plane, as we illustrate in  \figu{lgamma1} for \nehe\ to the WC detector and the Double Chooz best-fit $\stheta=0.08$. In this figure, sensitivity is given in the shaded regions to the corresponding performance indicators in the entire $\stheta$ range remaining after Double Chooz. The minimal possible $\gamma$, for which our conditions are fulfilled, is marked by the horizontal line. It is limited by the MH measurement from the left, and by the CPV measurement from the bottom. This means that the MH measurement leads to a sharp constraint $L \gtrsim L_{\mathrm{min}} \simeq 500 \, \mathrm{km}$, whereas the CPV measurement requires $\gamma \gtrsim 160$.  The figure illustrates what is characteristic for a large fraction of the parameter space: The baseline window for the minimal~$\gamma$ is rather sharp, and requires a fine-tuning of the detector location. Therefore, we focus on a set of longer, fixed baselines in the following, which allow for stable predictions. For some of these, the minimal~$\gamma$'s are illustrated by the arrows.

\begin{figure}[t]
\begin{center}
\includegraphics[width=10cm]{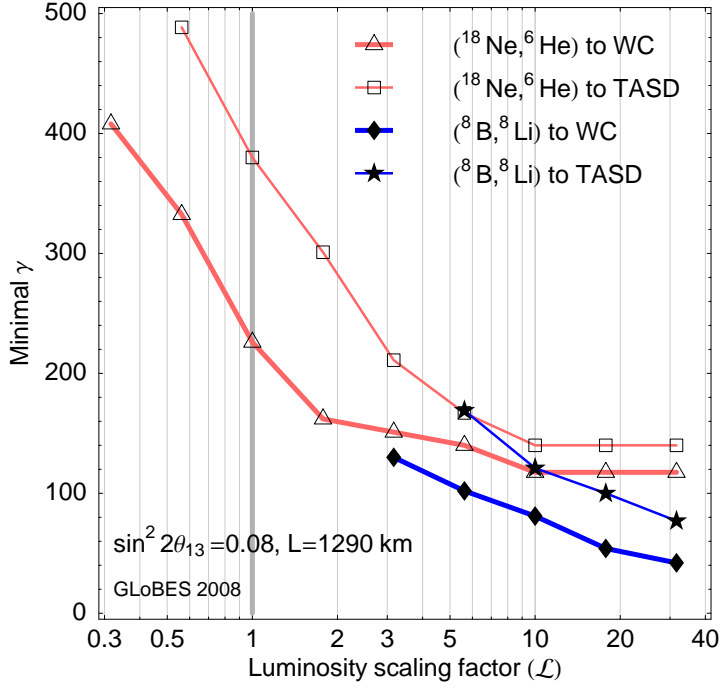}
\end{center}
\caption{\label{fig:lumiscale} Minimal $\gamma$ as a function of the luminosity scaling factor $\mathcal{L}$ for different isotope pair-detector combinations (in steps of 0.25 in $\mathrm{log}_{10} \mathcal{L}$). Here $L=1290 \, \mathrm{km}$ and the Double Chooz best-fit $\stheta=0.08$ from \Tab~\ref{tab:dchooz} are chosen.}
\end{figure}

In order to compare different detector technologies and isotope pairs as a function of $\mathcal{L}$, we show in \figu{lumiscale} the minimal~$\gamma$ for $L=1290 \, \mathrm{km}$ (fixed) and a Double Chooz best-fit $\stheta=0.08$ 
as an example. No symbol in this figure means that we have not found a setup which satisfies our criteria in the indicated $\gamma$ ranges.
Obviously, our chosen nominal luminosity $\mathcal{L}=1$ is sufficiently large for \nehe , but for \brli , $\mathcal{L} \gtrsim 5$ is required, \ie, \brli\ cannot be used at the nominal luminosity $\mathcal{L}=1$ to fulfill our requirements. We have tested that this conclusion holds irrespective of our discussed $\stheta$ case, detector, and baseline.
 We furthermore find that the WC detector outperforms the TASD because of the larger detector mass. As far as the different isotope pairs are concerned, the minimal possible $\gamma$ for \nehe\ becomes asymptotically limited for large $\mathcal{L}$ by the neutrino energies too low to allow for sufficient matter effects. This means that for large $\mathcal{L}$, the MH measurement limits the \nehe\ setups, whereas the \brli\ setups allow for a lower $\gamma$.

\begin{sidewaystable}[p]
\begin{center}
\begin{tabular}{l|rrrr|rrrr|rrrr}
\hline
 & \multicolumn{4}{c|}{$\stheta=0.04$} & \multicolumn{4}{c|}{$\stheta=0.08$} & \multicolumn{4}{c}{$\stheta=0.12$} \\
\bf{Setup} $\downarrow$ \hfill \bf{Baseline} [km] $\rightarrow$ &  ~730 & ~810 & 1050 & 1290 &   ~730 & ~810 & 1050 & 1290 & ~730 & ~810 & 1050 & 1290 \\ 
\hline
\bf{Beta beams} & & & & & & & & & & & &   \\
\nehe\ to WC, $\mathcal{L}=1$ & {\bf 220} & 230 & 290 & 350 & {\bf 200} & 210 & 240 & 230 & {\bf 190} & 200 & 220 & {\bf 190} \\
\nehe\ to TASD, $\mathcal{L}=1$ & - & {\bf 300} & 370 & 430 & {\bf 300} & 310 & 340 & 380 & {\bf 320} & {\bf 320} & 340 & 380 \\
\nehe\ to WC, $\mathcal{L}=5$ & {\bf 190} & {\bf 190} & {\bf 190} & 230 & {\bf 140} & {\bf 140} & {\bf 140} & {\bf 140} & {\bf 140} & {\bf 140} & {\bf 140} & {\bf 140} \\
\nehe\ to TASD, $\mathcal{L}=5$ & {\bf 200} & {\bf 200} & 220 & 230 & 180 & 180 & {\bf 170} & 180 & 180 & 170 & {\bf 160} & 170  \\
\brli\ to WC, $\mathcal{L}=5$ & - & - & {\bf 100} & 130 & {\bf 80} & {\bf 80} & 100 & 110 & {\bf 90} & {\bf 90} & 100 & 110 \\
\brli\ to TASD, $\mathcal{L}=5$ & - & - & {\bf 150} & 190 & - & - & {\bf 190} & {\bf 190} & - & - & - & {\bf 310} \\
\brli\ to WC, $\mathcal{L}=10$ & {\bf 70} & {\bf 70} & 90 & 110 & {\bf 60} & 70 & 80 & 90 & {\bf 60} & {\bf 60} & 70 & 80 \\
\brli\ to TASD, $\mathcal{L}=10$ & - & {\bf 100} & 130 & 140 & {\bf 110} & {\bf 110} & 120 & 130 & {\bf 120} & {\bf 120} & {\bf 120} & 130 \\
\hline
\bf{Superbeam upgrades}  & & & & & & & & & & & &\\
T2KK from \Ref~\cite{Barger:2007jq} & \multicolumn{4}{c|}{-} & \multicolumn{4}{c|}{$\surd$} & \multicolumn{4}{c}{$\surd$}\\
NO$\nu$A* from \Ref~\cite{Barger:2007jq} & \multicolumn{4}{c|}{-} & \multicolumn{4}{c|}{-} & \multicolumn{4}{c}{-}\\
WBB-120$_S$ from \Ref~\cite{Barger:2007jq} & \multicolumn{4}{c|}{-} & \multicolumn{4}{c|}{$\surd$} & \multicolumn{4}{c}{-}\\
\hline
\bf{Neutrino factories} & & & & & & & & & & & &\\
IDS-NF~1.0 from \Ref~\cite{ids} & \multicolumn{4}{c|}{$\surd$} & \multicolumn{4}{c|}{-} & \multicolumn{4}{c}{-}\\
Low-E NF from \Ref~\cite{Huber:2007uj} & \multicolumn{4}{c|}{-} & \multicolumn{4}{c|}{$\surd$} & \multicolumn{4}{c}{$\surd$}\\
\hline
\bf{Hybrids} & & & & & & & & & & & &\\
NF-SB from \Ref~\cite{Huber:2007uj} & \multicolumn{4}{c|}{$\surd$} & \multicolumn{4}{c|}{$\surd$} & \multicolumn{4}{c}{$\surd$}\\
\hline 
\end{tabular}
\end{center}
\caption{\label{tab:baseres} Minimal $\gamma$ (rounded up to the next 10) to measure all of the discussed performance indicators (see main text) at a specific baseline (in columns) for the given setups and Double Chooz $\stheta$ best-fit cases, where $\mathcal{L}$ is the luminosity scaling factor. In addition, a number of superbeam upgrades and neutrino factory setups are tested for the same criteria and same simulated values. A label ``-'' refers to no sensitivity in the discussed $\gamma$ ranges. The best options within each setup and $\stheta$ case are marked boldface. 
}
\end{sidewaystable}

We show in \Tab~\ref{tab:baseres} the minimal $\gamma$ (rounded up to the next 10) to measure all of the discussed performance indicators at specific baselines (in columns) for the given setups and Double Chooz $\stheta$ best-fit cases, where $\mathcal{L}$ is the luminosity scaling factor.  The different chosen baselines corresponds to CERN-LNGS or FNAL-Soudan (730~km), FNAL-Ash River (810~km), CERN-Boulby or JHF-Korea (1050~km), and FNAL-Homestake (1290~km).
Obviously, the minimal~$\gamma$ depends on the $\stheta$ case, which will be known after Double Chooz, and the baseline, which depends on the accelerator and detector locations. Therefore, once Double Chooz has found $\stheta$, one can easily read off this table the minimal~$\gamma$. If \brli\ can be used at reasonably high source luminosities ($\mathcal{L}=10$), $\gamma$ can be as low as about 60. If, however, \nehe\ is used at a lower luminosity, a $\gamma$ of at least 190 will be required. In addition, we show in \Tab~\ref{tab:baseres} a number of superbeam upgrades and neutrino factory setups are tested for the same criteria, same simulated values, and same $\stheta$ cases, where the details are given in the respective references. From this comparison, it is clear that almost none of the simulated alternatives can satisfy our criteria for any value of $\stheta$. However, if, for example, $\stheta=0.08$, T2KK, a wide-band beam (WBB-120$_S$, in this case using a 100~kt LArTPC), or a low energy neutrino factory might be used. The only setup in this list which can measure all of the performance indicators for all values of $\stheta$ is the NF-SB hybrid from \Ref~\cite{Huber:2007uj}. It combines a superbeam with a low energy neutrino factory beam directed towards the same detector in a distance of about $1 \, 250 \, \mathrm{km}$. From this comparison to alternative setups, it should be clear that the fraction of $\deltacp$ of 80\%, which we have initially used for CPV, is a good benchmark value on the edge of alternative setups.

\section{Summary and conclusions}

We have studied the minimal requirements for a single baseline beta beam experiments for large $\stheta$.
We have assumed that Double Chooz finds $\stheta$, and we have required that the next generation long-baseline
experiment measure mass hierarchy and CP violation at $3\sigma$ in the entire remaining $\stheta$ allowed region.
We have demonstrated that the minimal beta beam baseline is about $500 \, \mathrm{km}$. For any fixed baseline longer than this threshold, we have determined the minimal allowable $\gamma$.
Let us conclude depending on the geographical region, where our discussion is based on \Tab~\ref{tab:baseres}.
For Europe, the CERN-SPS might be used as an accelerator. The baseline to Frejus is not sufficient for a beta beam due to small matter effects. However, CERN-LNGS or CERN-Boulby can be used. If the SPS is not upgraded, \brli\ must be used at a high ion source luminosity $\mathcal{L} \gtrsim 5$, which might be achievable using a production ring.  If the SPS can be upgraded, \nehe\ at a lower ion source luminosity can be used as well. A $\gamma$ as low as about 200 could be sufficient for large $\stheta$ (for both ions), whereas a $\gamma$ as high as 350 might not be necessary~\cite{BurguetCastell:2005pa}. In addition, DESY might be used as a beta beam source, which opens new possibilities as long as $L \gtrsim 700 \, \mathrm{km}$. 
For the US, baselines such as FNAL-Soudan, FNAL-Ash River, or FNAL-Homestake are perfect for a beta beam experiment irrespective of the discussed $\stheta$ case. Since the Tevatron allows for higher $\gamma$'s than the SPS, \nehe\ might be used at our nominal ion source luminosity. For example, if a beta beam is directed towards a possibly existing large water Cherenkov detector at the Homestake mine, a $\gamma$ as low as 190 could be sufficient.
Compared to a wide band beam, which is limited by the proton intensity and target power, the $\gamma$ can be chosen high enough to allow for all measurements for any discussed $\stheta$ case.
For Japan, a baseline to Korea is perfectly suited for a beta beam, while the T2K baseline of 295~km is too short.
Compared to its alternatives, a beta beam might be the most flexible approach to measure all remaining quantities for large $\stheta$. For a given ion pair, source luminosity, and $\stheta$ case, we obtain a certain minimal~$\gamma$ which allows us to measure all remaining performance indicators to sufficient precisions. The resulting minimal~$\gamma$'s are not unrealistically high to outperform almost any alternative superbeam or neutrino factory setup. Therefore, there might be a clear neutrino oscillation physics case for the beta beam if $\stheta$ turns out to be large. In addition, synergies with nuclear physics applications may make a beta beam the most attractive alternative. For example, a low $\gamma$ beta beam (or an off-axis beta beam) could be used to obtain complementary information on neutrino-nucleus interactions, which might be even relevant for $0\nu\beta\beta$ experiments~\cite{Volpe:2003fi,Serreau:2004kx}.

{\bf Acknowledgments. }
I would like to acknowledge support from Emmy Noether program of Deutsche Forschungsgemeinschaft.

\end{document}